\documentclass[]{article}

\usepackage{mathptmx}       % selects Times Roman as basic font
\usepackage{helvet}         % selects Helvetica as sans-serif font
\usepackage{courier}        % selects Courier as typewriter font
\usepackage{type1cm}        % activate if the above 3 fonts are
\usepackage{makeidx}         % allows index generation
\usepackage{graphicx}        % standard LaTeX graphics tool
\usepackage{multicol}        % used for the two-column index
\usepackage[bottom]{footmisc}% places footnotes at page bottom

\usepackage{glossaries}
\include{gloss}
\makeglossaries
\makeindex             % used for the subject index
\usepackage{array}

\usepackage{subfigure}
\usepackage{latexsym}
\usepackage{verbatim}
\usepackage{epsfig}
\usepackage{amsmath}
\usepackage{epstopdf}
\usepackage{times}
\usepackage{amsfonts}
\usepackage{graphics}
\usepackage{url}
\usepackage{amssymb}

\newcommand{\score}{{{\em score}}}
\newcommand{\cscore}{{{\cal {CS}}}}
\newcommand{\mean}{{{\cal \bold M}}}

\begin{document}

\title{Overlapping Community Discovery \\ Methods:  A Survey}
					% Activate to display a given date or no date

\author{Alessia Amelio and  Clara Pizzuti\\
National Research Council of Italy (CNR)\\
Institute for High Performance Computing and Networking (ICAR), \\
Via Pietro Bucci, 41C\\
87036 Rende (CS), Italy\\
email:\{amelio,pizzuti\}@icar.cnr.it}

\date{}

\maketitle

\abstract{The detection of overlapping communities is a challenging problem which is gaining increasing interest in recent years because of the natural attitude of individuals, observed in real-world networks,  to participate in multiple groups at the same time. This review gives a description of the main proposals in the field. Besides the methods designed for static networks, some new approaches that deal with the detection of overlapping communities in networks that change over time, are described. Methods are classified with respect to the underlying principles guiding them to obtain a network division in groups sharing part of their nodes. For each of them we also report, when available,  computational complexity and web site address from which it is possible to download the software implementing the method.}

%\section{}
%\subsection{}

\section{Introduction}
Complex networks constitute an efficacious formalism to represent
the relationships among objects composing many real world
systems. Collaboration networks, the Internet, the world-wide-web,
biological networks, communication and transport networks, social
networks are just some examples. Networks are modeled as graphs,
where nodes represent objects and edges represent
interactions among these objects. One of the main problems in the
study of complex networks is the detection of {\it community
structure}, i.e. the division of a network into groups (clusters
or modules) of nodes having dense intra-connections, and sparse
inter-connections. In the last few years many different approaches
have been proposed to uncover community structure in networks. 

Much of the effort in defining efficient and efficacious methods for community detection has been directed on finding disjoint communities. 
However, communities may overlap, i.e. some nodes may belong to
more than one group. The membership of an entity to many groups
is very common in real world networks. For example, in a social
network, a person may participate in several interest groups, in collaboration networks a researcher may collaborate with many groups, in citation networks a paper could involve more than one topic, in biological networks proteins play
different roles in the cell by taking part in several processes. 

In this review a description of  the most recent proposals for overlapping community detection is given, and a classification in different categories is provided. Algorithms have been classified by taking into account the underlying principles guiding the methods to obtain network division in groups sharing part of their nodes.  

Many recent reviews describing community detection algorithms have been published \cite{Fortunato2007,Lancichinetti2009,Fortunato2010,Coscia2011,Xie2013}. However, our review differs from those of  \cite{Fortunato2007,Lancichinetti2009,Fortunato2010,Coscia2011} since we focus only on overlapping approaches, and from \cite{Xie2013} because also dynamic approaches are described. 

The paper is organized as follows.  The next section gives some preliminary definitions necessary for the description of the approaches. Section \ref{metodi} introduces the method categorization proposed in this review. Section \ref{nsle} describes \emph{node seeds and local expansion methods}. Section \ref{ce} considers \emph{clique expansion methods}. Section \ref{lc} describes \emph{link clustering algorithms}. Section \ref{lp} presents \emph{label propagation approaches}.  Methods for which a categorization in one of the defined classes was not possible are reported in Section \ref{oa}. Section \ref{dn} considers the more recent proposals for dynamic networks. Section \ref{bench} gives some information regarding benchmarks that can be used to test algorithm performance. Finally, Section \ref{concl} gives some final considerations on the described methods and concludes the paper.   

\section{Preliminaries}
In this section some basic definitions, necessary for a clear understanding of the concepts described in the survey, are given. 

A network $\cal N$ can be modeled as  a graph $G=(V,E)$ where $V$
is a set of $n=\mid V\mid$ objects, called nodes or vertices, and $E$ is a set of
$m=\mid E \mid$ links, called edges,  that connect two elements of $V$. 
A community  in a  network is a group
of vertices (i.e. a subgraph) having a high density of edges
among them, and a lower density of edges between groups. In \cite{Fortunato2010} it is observed that a formal definition of community does not exist because this definition often depends on the application domain. Nevertheless, an impressive number of methods has been proposed to detect communities in complex networks. Before starting with the description of overlapping approaches, it is necessary to introduce the concept of $modularity$ because of its popularity and large use among researchers. 

The concept of $modularity$ has  been originally defined by Girvan and Newman \cite{NewmanGirvan2004}  as quality function to evaluate the goodness of a partition. Along the years, however, it has been accepted as one of the most meaningful measures to partition a network, that more closely agrees with the intuitive concept of community on a wide range of real world networks.

The idea underlying modularity is that a  {random graph} has not a clustering structure, thus the edge density of a cluster should be higher than the expected density of a subgraph whose nodes are connected at random. This expected edge density depends on a chosen \emph{null model}. Modularity can be written in the following way: 

\begin{equation}Q=  \frac{1}{2m} \sum_{ij}(A_{ij} -P_{ij})\delta(C_i, C_j) \end{equation} \label{q0}

\noindent where $A$ is the adjacency matrix of the graph $G$, $m$ is the number of edges of $G$, and $P_{ij}$ is the expected number of edges between nodes $i$ and $j$ in the null model. $\delta$ is the Kronecker function and yields one if $i$ and $j$ are in the same community, zero otherwise. When it is assumed that the random graph has the same  {degree distribution} of the original graph,  $P_{ij} = \frac{k_i k_j}{2m}$, where $k_i$ and $k_j$ are the  {degrees of nodes} $i$ and $j$ respectively. Thus the modularity expression becomes:

\begin{equation} Q=  \frac{1}{2m} \sum_{ij}(A_{ij} - \frac{k_i k_j}{2m})\delta(C_i, C_j)  \end{equation} \label{q1}

\noindent Since only the pairs of vertices belonging to the same cluster contribute to the sum,  modularity can be rewritten as    

\begin{equation} Q= \sum_{s=1}^{k} [\frac{l_s}{m} - ( \frac{d_s}{2m})^2]
 \end{equation} \label{q2}

\noindent where $k$ is the number of modules found inside a network, 
 $l_s$ is the total number of edges joining vertices inside
the module $s$, and $d_s$ is the sum of the degrees of the nodes
of $s$. Thus the first term of each summand is
the fraction of edges inside a community, and the second one is the
expected value of the fraction of edges that would be in the
network if edges fell at random without regard to the community
structure. Values approaching 1 indicate strong community
structure.    

\section{Methods}\label{metodi}
In this section, methods for the detection of overlapping communities are described. They have been classified in six different categories on the base of the methodology employed to identify communities. The categories are the following:
\begin{itemize}
\item Node seeds and local expansion
\item Clique expansion
\item Link clustering
\item Label propagation
\item Other approaches
\item Dynamic networks
\end{itemize}

For each category, a short description of the main common features among algorithms of that class is provided.

\subsection{Node Seeds and Local Expansion} \label{nsle} 
The idea underlying these approaches is that starting from a node or a small set of nodes, a community can be obtained by adding neighboring nodes that improve a quality function. The quality function characterizes the structure of the obtained clustering.  

Baumes et al.  \cite{Baumes2005a}  introduced the concept of density function and defined a community as a subgraph that is locally optimal with respect to this density function. The internal $p_{in}$ and external $p_{ex}$ edge intensities of a community $C$ are defined as 

 \begin{equation} p_{in}(C)=\frac{E(C)}{\mid C \mid (\mid C \mid -1)} ~~~~~~~ p_{ex}(C)= \frac{E_{out}(C)}{2 \mid C \mid (n - \mid C \mid)}
 \end{equation} \label{pin}
 
\noindent  where $E(C)$ is  the number of internal edges of $C$ and $E_{out}(C)$ is the number of edges from nodes of $C$ towards nodes not belonging to $C$. 
 Three weight or metric functions are defined to assign a weight to a graph. The \emph{internal edge probability} 
 \begin{equation}W_p =  p_{in}(C)\end{equation}\label{wp}
 
 \noindent that coincides with the internal edge density, the \emph{edge ratio}  
 \begin{equation}W_e= \frac{E(C)}{E(C) + E_{out}(C)}\end{equation}\label{we} 
 
 \noindent and the \emph{intensity ratio} 
 \begin{equation}W_i= \frac{p_{in}(C)}{p_{in}(C) + p_{ex}(C)}
 \end{equation}\label{wi}
 
\noindent These metrics measure the intensity of communication within the clusters, and can be efficiently updated when a new node is added or removed from the cluster. Finally, to measure the difference between two clusters, the  {Hamming}, or  {edit distance}, and the percentage of non-overlap are defined. 
 
In order to find overlapped communities, the authors proposed two methods.  The first algorithm,  \emph{Iterative Scan (IS)}, starts by choosing an edge at random, called $seed$, and adds or removes one vertex at a time until the chosen density metric  improves. When no more improvement can be obtained, the algorithm stops and it is restarted with a new seed. Overlapping is possible because the restart process can reassign nodes already present in a cluster to another new forming community.

 The second algorithm, \emph{Rank Removal (RaRe)}, assumes that there are some high-ranking nodes which, when removed from the graph, disconnect the graph into smaller connected components, called cores. The deleted nodes are then added to one or more cores. This means that the overlapping between two clusters is possible only through these vertices.  

To validate the methods, the Hamming distance between each  cluster of the true clustering and the  obtained clustering is computed, then the average of all these distances is considered.

The choice of a random edge can negatively influence the result of $IS$. Thus the same authors modified their \emph{Iterative Scan} method  and proposed a more efficient algorithm  for finding overlapping communities \cite{Baumes2005b}, named $IS^2$, that combines $IS$ and $RaRe$. $IS^2$ also relies on a new strategy for initializing seed clusters able to compute the ranking of each node only once. The strategy is based on the observation that the only nodes capable of increasing cluster density are either the members of the cluster itself or members of the immediate neighborhood clusters, where neighboring clusters are those containing nodes adjacent to a node inside the cluster. Thus, rather than visiting each node for each iteration, nodes not belonging to one of these two groups can be skipped over.

Another variation of the $IS$ method, that ensures connectivity of the detected communities, is described in  \cite{Goldberg2010}. \\

Lancichinetti et al.  \cite{Lancichinetti2008a} proposed a method to detect overlapping and hierarchical community structure, in the following referred as $LFM$,  based on the concept  of \emph{fitness}  of nodes belonging to a community $S$. Let   $k_i^{in}(S)$ and $k_i^{out}(S)$ be the internal and external degrees of the nodes belonging to a community $S$. The \emph{fitness}  of $S$ is then defined as 
\begin{equation} {\cal F}_{S} = \sum_{i\in S } \frac{k_i^{in}(S) } {(k_i^{in}(S) + k_i^{out}(S))^{\alpha}}
\end{equation}\label{fs}

\noindent where  $\alpha$, called resolution parameter, is a positive real-valued parameter controlling the size of the communities. When $ k_i^{out}(S) = 0 ~~ \forall  i$, ${\cal F}_{S}$ reaches its maximum value for a fixed $\alpha$.
The community fitness has been used in \cite{Lancichinetti2008a}  to find communities one at a time.  \emph{Node fitness} with respect to a community $S$ is defined as the variation of  \emph{community fitness}  of $S$ with and without the node $i$, i. e.
\begin{equation} {\cal {F}}^i_{S} = {\cal {F}}_{S \cup  \{i\}} - {\cal {F}}_{S -  \{i\}} 
\end{equation}\label{fis}

\noindent The method starts by picking a node at random, and considering it as a
community $S$. Then a loop over all the neighbor nodes of $S$, not included in $S$, is
performed in order to choose the neighbor node to be added to $S$.
The choice is done by computing the node fitness  for each node,
and augmenting $S$ with the node having the highest value of fitness.
At this point the fitness of each node is recomputed, and
if a node turns out to have a negative fitness value it is removed from
$S$. The process stops when all the not yet included neighboring nodes of the nodes in  $S$  have a negative fitness. Once a community has been obtained, a new node is picked
and the process restarts until all the nodes have been assigned to
at least one group. The authors found that the divisions obtained for the resolution parameter $\alpha=1$ are relevant. However, they introduce a criterion to choose a clustering based on the concept of stability. A clustering is considered stable if it is delivered  for a range of values of $\alpha$. The length of this range determines the more stable division, which is deemed the best result. 

A different approach is applied by Lancichinetti et al. in \cite{Lancichinetti2011}  to choose the neighbor of a node  to add to a cluster $S$. In fact, a method called $OSLOM$ uses a statistical test to grow a community $S$ by evaluating the statistical significance of a node. The intuition under the approach is that, if a vertex $v$  shares many more edges with the nodes of $S$  than expected in the null model, then  the relationship between $v$ and $S$  is ÒunexpectedlyÓ strong, thus $v$ can be included  in $S$. \\

\emph{DOCS (Detecting Overlapping Community Structures)} \cite{Wei2009} is a method based on an approach of global division followed by local expansion. $DOCS$ uses classical spectral graph partitioning and  {random walk} techniques. It first applies a spectral bi-section method with multi-level recursion to generate seed groups, and then employs a process of local expansion to add a vertex to the current cluster. 
The local optimization process randomly walks over the network from the seeds. In particular, the locally-optimal expansion process is based on the concept of modularity $Q$. At each time step, the scanned vertices are sorted in descending order with the degree normalized probabilities. If a vertex brings better $Q$ change to the community candidates, it may be absorbed as a new member of the community structure. Furthermore, the algorithm, besides trying  to achieve a good $Q$ value,  takes also into account the overlapping rate of the discovered communities. The user must fix a threshold $t$, and the expansion is continued until the overlapping rate is beyond the $t$ value. \\

$Moses$ \cite{McD2010} is a  {greedy algorithm} that optimizes a global objective function based on a statistical network model.  In this model a graph is represented by a random symmetric adjacency matrix.  The objective function computes the maximum likelihood estimators from the observed likelihood. $Moses$ selects an edge ($u,v$) at random and the community $C=\{u,v\}$ constituted by the two corresponding nodes  is built by expanding it with new nodes taken from the set of neighboring nodes not yet added to  $C$.  Nodes are selected such that the objective function is maximized, and expansion continues until the highest value of the objective is obtained. Since edges are chosen at random with replacement, the same edge could be selected many times, and expanded in different communities. The algorithm periodically checks for all the communities found so far if the removal of one community improves the objective function. Furthermore, a tuning phase at the end of the expansion is performed. In this phase each node is removed from all the communities it has been assigned, and added to the communities to which it is connected by an edge. The change is accomplished only if the objective function increases.   $Moses$ has been compared with the methods $LFM$ of Lancichinetti et. al. \cite{Lancichinetti2009}, $COPRA$  of Gregory \cite{Gregory2009}, \emph{Iterative Scan} ($IS$) of Baumes et al. \cite{Baumes2005a}, $GCE$ of Lee et al. \cite{Lee2010}, $CFinder$ of Palla et al. \cite{Palla05} . The authors performed experiments by varying the overlap to range from one to ten communities per node, i.e. a node can participate from one until 10 communities. They found that when the average overlap increases, $Moses$ is able to recover community structure better than the other methods.

\subsection{Clique expansion} \label{ce} 
Clique expansion methods are similar to the approaches described in the previous section. However, they consider as seed cores highly connected sets of nodes constituted by  {cliques}, and then generate overlapping communities by merging these cliques, applying different criteria. \\

$CFinder$ \cite{Palla06} is a system to identify and visualize overlapped, densely
connected groups of nodes in undirected graphs. It also allows to
navigate the original graph and the communities found. The search
algorithm $CFinder$ uses the \emph{Clique Percolation Method}
\cite{Palla05} to find $k$-$clique$ percolation clusters. A
k-clique is a complete subgraph constituted by $k$ nodes. Two
cliques are said adjacent if they share exactly $k-1$ nodes. A
k-clique  percolation cluster is defined as the union of all
k-cliques that can be reached from each other through a series of
adjacent k-cliques.  The parameter $k$ has to be provided in
input. The higher the value of $k$, the smaller the size of the
highly dense groups. The authors suggest that a value between 4
and 6 gives the richest group structure. $CFinder$ works as follows.
First of all, the community finding algorithm extracts all the
maximal complete subgraphs, i.e. the cliques. Then a clique-clique
overlap matrix is prepared. In this matrix each row (and column)
corresponds to a clique, the diagonal contains the size of the
clique, while the value contained in the position ($i,j$) is the
number of common nodes between the cliques $i$ and $j$. 
K-clique communities are obtained by deleting every element on the
diagonal smaller than $k$, and every element off the diagonal smaller
than $k-1$, replacing the remaining elements by one, and finally
finding the connected components in the modified matrix. \\

Shen et al. \cite{Shen2009}  proposed a hierarchical agglomerative algorithm named $EAGLE$ (\emph{agglomerativE HierarchicAL clusterinG based on maximaL cliquE}) to uncover hierarchical and overlapping community structure in networks. The method is based on the concepts of \emph{maximal cliques}, i.e. cliques that are not a subset of another clique, and \emph{subordinate maximal cliques}, i.e. maximal cliques whose vertices are contained in some other larger maximal clique. Fixed a threshold $k$, all the subordinate maximal cliques with size smaller than $k$ are discarded. The deletion of these cliques implies that some vertices do not belong to any maximal clique. These vertices are called \emph{subordinate vertices}. The algorithm starts by considering as initial communities the maximal cliques and the subordinate vertices. Then, for each couple of communities, the similarity between them is computed and the two groups with maximum similarity are merged. This process is repeated until  only one community remains.  The similarity $M$ between two communities is defined by specializing the concept of modularity, introduced by Newmann, for two communities $C_1$ and $C_2$:
\begin{equation} M= \frac{1}{2m} \sum_{v \in C_1, w \in C_2 , v\neq w} [ A_{vw} -\frac{k_vk_w}{2m}]
\end{equation}\label{m}

\noindent where $m$ is the total number of edges in the network, $A$ is the adjacency matrix of the network, $k_v$ is the degree of node $v$. The evaluation of the results obtained is performed by computing an extended modularity value $EQ$ that takes into account the number of communities a node belongs to. This value is computed as follows:
\begin{equation} EQ= \frac{1}{2m} \sum_i \sum_{v \in C_i, w \in C_i }\frac{1}{O_v O_w} [ A_{vw} -\frac{k_vk_w}{2m}]
\end{equation}\label{eq}

\noindent where $O_v$ is the number of communities to which $v$ belongs. The value of $EQ$ is used by the authors to cut the generated  {dendrogram}  and to select the community structure having the maximum extended modularity value. Experiments on two real life networks show that the results obtained are meaningful.\\

\emph{Greedy Clique Expansion} ($GCE$) is a community detection algorithm proposed by Lee at al. \cite{Lee2010} that assigns nodes to multiple groups by expanding cliques of small size. A clique is considered as the seed or core of a community $C$, and it is used as starting point to obtain $C$ by greedily adding nodes that maximize a fitness function. The fitness function adopted is that proposed by Lancichinetti et al. \cite{Lancichinetti2009}. The algorithm first finds the maximal cliques contained in the graph $G$ representing the network, with at least $k$ nodes. Then it creates a candidate community $C'$ by selecting the largest unexpanded seed and adding the nodes contained in the frontier that maximize the fitness function. This expansion is continued until the inclusion of any node would lower the fitness. At this point $GCE$ checks if the community $C'$ is near-duplicate, i.e. $C'$ is compared with all the already obtained communities, and it is accepted only if it is sufficiently different from these communities. All the steps are repeated until no seeds remain to be considered. In order to decide if a community is near-duplicate, a distance measure between communities, based on the percentage of uncommon nodes, is introduced. This measure is defined as follows. Given two communities $S$ and $S'$, 

\begin{equation}\delta_E(S,S')=1 - \frac{\mid S\cap S' \mid}{min(\mid S \mid,\mid S' \mid)}\end{equation}\label{deltae}

\noindent This measure can be interpreted as the proportion of nodes belonging to the smaller community that are not members of larger community. Fixed a parameter $\epsilon$ as the minimum community distance, and a set of communities $W$, the near-duplicates of $S$ are all the communities in $W$ that are within a distance $\epsilon$ from $S$.

$GCE$ has been compared with other state-of-the-art overlapping community detection methods, and, analogously to $Moses$, it performs better when the number of communities that a node can belong to increases from 1 to 5.  

\subsection{Link Clustering} \label{lc} 
Link clustering methods propose to detect overlapping communities by partitioning the set of links rather than the set of nodes. To this end, the \emph{line graph} is used. The {\it line graph} $L(G)$ of an undirected graph $G$ is
another graph $L(G)$ such that 1) each vertex of $L(G)$ represents
an edge of $G$, and 2) two vertices of $L(G)$ are adjacent if and
only if their corresponding edges share a common endpoint in $G$.
Thus a line graph represents the adjacency  between edges of $G$.
The main advantage of applying clustering to the line graph is that it produces an overlapping graph division of the original interaction graph, thus allowing
nodes to be present in multiple communities. 

Pereira et al. \cite{PE04} have been the first in using the line graph to find overlapping modules for protein-protein interaction networks. To this end they applied a well known method  (MCL \cite{EDO02}) for protein interaction networks to the line graph.\\

Evans and Lambiotte \cite{Evans2009} argued that any algorithm that partitions a network can be applied to partition links for discovering overlapping community structure. They first reviewed a definition of modularity which uses statistical properties of dynamical processes taking place on the edges of a graph, and then proposed link partitioning by applying the new modularity concept.  In particular, the traditional modularity $Q$ is defined in terms of a random walker moving on the links of the network. Such a walker would therefore be located on the links instead of the nodes at each time $t$, and its movements are between adjacent edges, i.e. links having one node in common. Three quality functions %$Q(C)$, $Q(D)$ and $Q(E1)$ 
for partitioning links of a network $G$ have been proposed. Each formalizes a different dynamical process and explores the structure of the original graph $G$ in a different way. 

In the first dynamical process,  \emph{Link-Link random walk}, the walker jumps to any of the adjacent edges with equal probability. In the second process, \emph{Link-Node-Link random walk}, the walker moves first to a neighboring node with equal probability, and then jumps to a new link, chosen with equal probability from those new edges incident at the node. In the last process, the dynamics are driven by the original random walk but are projected on the links of the network. The stabilities of the three processes have been defined by generalizing the concept of modularity to paths of arbitrary length, in order to tune the resolution of the optimal partitions. The optimal partitions of these quality functions can be discovered by applying standard modularity optimization algorithms to the corresponding line graphs.
An extension to deal with weighted line graphs is reported in \cite{Evans2010}.\\

\emph{GA-NET+} is a method proposed by Pizzuti \cite{Pizzuti2009}  that employs  \emph{Genetic Algorithms} \cite{goldberg89}. 
A Genetic Algorithm ($GA$) evolves a constant-size population of 
elements (called $chromosomes$) by using the genetic operators of {\it reproduction}, {\it crossover} and {\it mutation}. Each chromosome represents a candidate solution to a given problem and it is associated with a {\it fitness value} that reflects how good it is, with respect to the other solutions in the population. The method uses the concept of \emph{community score} to measure the quality of the division in communities of a network. Community score is defined as follows.

Let $\mu_{i}$ denote the fraction of edges connecting node $i$ to
the other nodes in a community $S$. More formally
\begin{center}
$\mu_{i} = \frac{1}{\mid S \mid} k_i^{in}(S) $
\end{center}

\noindent where $\mid S \mid$ is the cardinality of $S$, and $k_i^{in}(S) = \sum_{j\in
S} A_{ij}$ is the number of edges connecting $i$ to the other
nodes in $S$.

The {\it power mean of $S$ of order} $r$, denoted as $\mean(S)$ is
defined as

\begin{equation} \mean(S)~ =   \frac{\sum_{i\in S} (\mu_{i})^r}{|S|} 
\end{equation}\label{means}

\noindent The {\it volume} $v_{S}$ of a community $S$ is defined as the
number of edges connecting vertices inside $S$, i.e the number of
$1$ entries in the adjacency sub-matrix of $A$ corresponding to
$S$, $v_{S} = \sum_{i,j\in S }A_{ij}$.

The \score~ of $S$ is defined as $score(S) = \mean(S) \times
v_{S}$.  The {\it community score} of a clustering $\{S_1,
\ldots S_k\}$ of a network  is defined as
\begin{equation} \cscore = \sum^k_i score(S_i)
\end{equation}\label{cscore}

\noindent {\it Community score} gives a global measure of the
network division in communities by summing up the local score of
each module found. The problem of community identification can
then be formulated as the problem of maximizing $\cscore$.

The algorithm tries to maximize $\cscore$ by running the genetic algorithm on the line graph $L(G)$ of the graph $G$ modeling the network. As already pointed out,  a main advantage in using the line graph is that the partitioning of $L(G)$ obtained by \emph{GA-NET+} corresponds to an overlapping graph division of $G$. 
The method uses the locus-based adjacency representation. In this representation an individual of
the population consists of $N$ genes $g_1,\ldots,g_N$ and each
gene can assume allele values $j$ in the range $\{1, \ldots, N\}$.
Genes and alleles represent nodes of the graph $G=(V,E)$ modelling
a   network $\cal N$, and a value $j$ assigned to the $i$th
gene is interpreted as a link between the nodes $i$ and $j$ of
$V$, thus in the clustering solution found $i$ and $j$
will be in the same cluster.  \emph{GA-NET+}  starts by generating a population initialized at random with individuals representing a partition in subgraphs of the line graph $L(G)$. After that, the fitness of the individuals from the original graph must be evaluated and a new population of individuals is created by applying uniform crossover and mutation. The dense communities present in the network structure are obtained at the end of the algorithm, without the need to know in advance the exact number of groups.  This number is automatically determined by the optimal value of the community score.\\

Ahn et al.  \cite{Ahn2010} proposed a hierarchical agglomerative link clustering method to group links into topologically related clusters. The algorithm applies a hierarchical method to the line graph by defining two concepts:  \emph{link similarity} and  \emph{partition density}. Link similarity is used during the  {single-linkage} hierarchical method to find the pair of links with the largest similarity in order to merge their respective communities. This similarity measure is defined as follows. Let a node $i$ be given, the \emph{inclusive neighbors} of $i$ are 
\begin{equation}n_+(i)=\{x \mid d(i,x) \leq 1 \}\end{equation}\label{ni}

\noindent where $d(i,x)$ is the length of the shortest path between nodes $i$ and $x$. Thus $n_+(i)$ contains the node itself and its neighbors. Then, the similarity $S$ between two links $e_{ik}$ and $e_{jk}$ can be defined by using the Jaccard index:
\begin{equation}S(e_{ik},e_{jk})=\frac{\mid n_+(i) \cap n_+(j) \mid}{\mid n_+(i) \cup n_+(j) \mid}
\end{equation}\label{see}

\noindent The similarity between links is also extended to networks with weighted, directed, or signed links (without self-loops).
The agglomerative process is repeated until all links belong to a single cluster. To find a meaningful community structure, it is necessary to decide where the built dendrogram must be cut. To this end, the authors introduce a new quantity, the \emph{partition density D}, that measures the quality of a link partitioning. 

Partition density is defined as follows. Let $m$ be the number of links of a given network, and $\{ P_1, \ldots, P_C\}$ the partition of the links in $C$ subsets. Each subset $P_c$ has $m_c = \mid P_c \mid$ links and $n_c = \mid \cup_{e_{ij} \in P_c} \{i,j\} \mid$ nodes.  Then 

\begin{equation}D_c = \frac{m_c - (n_c -1)}{n_c (n_c -1)/2 - (n_c -1)} = 2 \frac{m_c - (n_c -1)}{(n_c - 2) (n_c -1)}
\end{equation}\label{dc}

\noindent is the normalization of the number of links $m_c$ by the minimum and maximum number of possible links between $n_c$ connected nodes. It is assumed that $D_c =0$ when $n_c =2$. The partition density $D$ is the average of the $D_c$, weighted by the fraction of present links: 
 \begin{equation} D= \frac{2}{m} \sum_c m_c \frac{m_c - (n_c -1)}{(n_c - 2)(n_c -1)}
 \end{equation}\label{d}

\noindent The authors, in order to compare their approach with other state-of-the-art methods, introduce four measures. Two measures, \emph{community quality} and \emph{overlap quality}, are based on  metadata known possessed by some networks studied in the literature. These metadata consist of a small set of annotations or tags attached to each node. The other two measures, \emph{community coverage} and \emph{overlap coverage}, consider the amount of information extracted from the network. 
Community coverage counts the fraction of nodes that belong to at least one community of three or more nodes, called nontrivial communities. The authors state that this measure provides a sense of how much of the network is analyzed. Overlap coverage counts the average number of membership per nodes to nontrivial communities. When the communities are not overlapping, the two coverage measures give the same information. 

\subsection{Label propagation} \label{lp} 
In label propagation approaches a community is considered a set of nodes which are grouped together by the propagation of the same property, action or information in the network.\\

%\noindent
%\emph{Gregory  \cite{Gregory2010}}

Gregory in \cite{Gregory2010} proposed the algorithm $COPRA$ (\emph{Community Overlap PRopagation Algorithm}), as an extension of the label propagation technique of Raghavan et al. \cite{Raghavan2007}. The main modification consists in assigning multiple community identifiers to each vertex. The method, thus, associates with each vertex $x$ a set of couples ($c,b$), where $c$ is a community identifier and $b$ is a belonging coefficient expressing the strength of $x$ as member of community $c$. $COPRA$ starts by giving to each vertex  a single label with belonging coefficient set to 1. Then, repeatedly, each vertex $x$ updates its labels by summing and normalizing the belonging coefficients of its neighboring nodes. The new set of $x$'s labels is constituted by the union of its neighbor labels. However, in order to limit the number of communities a vertex can participate, a parameter $v$ must be given in input.  In particular, the labels whose belonging coefficient is less that $1/v$ are deleted. $COPRA$ has a nondeterministic behavior when all the belonging coefficients corresponding to the labels associated with a vertex are the same, but below the threshold. In such a case a randomly selected label is maintained, while the remaining are discarded. 
Finally, communities totally contained in others are removed, and disconnected communities that could be generated are split in connected ones. \\

Wu et al. in \cite{Wu2012} pointed out that the input parameter $v$ makes  $COPRA$ unstable since it is a global vertex-independent parameter  not taking into account that, often, most nodes are non-overlapping, while few nodes participate in many communities.  Thus an appropriate choice of $v$ is difficult and it induces the non-determinism described above. To overcome this shortcoming, Wu et al. proposed $BMLPA$ (\emph{Balanced Multi-Label Propagation Algorithm}), a method based on  a new label update strategy that computes balanced belonging coefficients, and does not limit the number of communities a node can belong to.  A balanced belonging coefficient is computed by normalizing each coefficient by the maximum value a vertex has, and retaining it only if its normalized value is above a fixed threshold $p$. Another characteristic introduced by $BMLPA$ is the initialization process of vertex labels based on the extraction of overlapping rough cores. Such cores allow $BMLPA$ to efficiently assign labels to each node, and to effectively update labels since the threshold $p$, independently the value chosen, would make the new update strategy not work well because each node would retain all the labels of its neighbors.    \\  

$SLPA$ \cite{Xie2011,Xie2012} is another extension of the label propagation technique of Raghavan et al. \cite{Raghavan2007} that adopts a speaker-listener based information propagation process. Each node is endowed with a memory to store the labels received. It  can have both the role of $listener$ and $speaker$. In the former case it takes labels from the neighbors and accepts only one following a listening rule, such as the most popular observed at the current step. If it is a speaker, it sends a label to the neighboring listener node by choosing a label with respect to a certain speaker rule, such as single out a label with probability proportional to its frequency in the memory. The algorithm stops when a fixed number $t$ of iterations has been reached.

\subsection{Other approaches} \label{oa} 
In this section methods which could not be categorized in one of the above classes are reported. \\

Zhang et al.  \cite{Zhang2007} developed an algorithm for detecting overlapping community structure by combining modularity concept, spectral relaxation and fuzzy c-means. 
In particular, a new modularity function extending the Newman's modularity concept is introduced to take into account soft assignments of nodes to communities. The problem of maximizing the modularity function is reformulated as an eigenvector problem. Fixed an upper bound to the number $k$ of communities, the top $k-1$ eigenvectors of a generalized eigen system are computed, and a mapping of the network nodes into a $d$-dimensional Euclidean space is performed, where $d\leq k-1$. After that, fuzzy c-means clustering is applied to group nodes by maximizing the modified modularity function.\\

In \cite{Gregory2007} Gregory presented a hierarchical, divisive approach, based on Girvan and Newman's algorithm (GN) \cite{Girvan02}, but extended with a novel method of splitting vertices. $CONGA$ (\emph{Cluster-Overlap Newman Girvan Algorithm} ) adds to the GN algorithm the possibility to split vertices between communities, based on the concept of \emph{split betweenness}. This concept allows to choose either to split a vertex or remove an edge. The edge betweenness of an edge $e$ is the number of shortest paths, between all pairs of vertices, that pass along $e$. A high betweenness indicates that the edge acts as a bottleneck between a large number of vertex pairs and suggests that it is an inter-cluster edge. The split betweenness of a vertex $v$ is the number of shortest paths that would pass between the two parts of $v$ if it were split. A vertex can be split  in many ways; the best split is the one that maximizes the split betweenness. An approximate, efficient algorithm has been presented for computing split betweenness and edge betweenness at the same time. In $CONGA$, a network is initially considered as a single community, assuming it is connected. After one or more iterations, the network is subdivided into two components (communities). Communities are repeatedly split into two until only singleton communities remain. If binary splits are represented as a dendrogram, the network can be divided into any desired number of communities.

$CONGA$ has a complexity of $O(m^3)$, where $m$ is the number of edges, thus it is rather inefficient. A faster implementation of $CONGA$, named $CONGO$, that uses local betweenness and runs in $O(m~log~m)$ is proposed by the same author in \cite{Gregory2008}.\\

Chen et al. in \cite{Chen2010} proposed an approach  to find communities with overlaps and outlier nodes based on visual data mining. They consider a community as a network partition whose entities share some common features and a relationship metric is adopted to evaluate the proximity of the entities each other. 
Such metric is based on the notion of random connections useful to identify communities which are considered as non-random structures. Furthermore, it takes into account the neighborhood around any two nodes in order to evaluate their relationship. An algorithm that generates an ordering of the network nodes according to their relation scores is presented. From this ordered list of nodes, communities are obtained by considering a consecutive group of nodes with high relation score. 
A 2D  visualization of these scores shows peaks and valleys, where a sharp drop  of relation scores after a peak is interpreted as the end of a community, while the valleys between two peaks represent a set of hubs which belong to several communities. By this visualization a user is requested to fix a \emph{community threshold} and an \emph{outlier threshold} that allow the algorithm to decide whether a node should be considered an outlier or be added to the current community. The authors state that the main advantage of this visual mining approach is that a user can easily provide input parameters that allow the method to find communities, hub nodes, and outliers.\\

Rees and Gallagher \cite{Rees2010} proposed an approach to discover communities based on the collective viewpoint of individuals. The base concept is that each node in the network knows, by way of its \emph{egonet}, the members of its friendship group. An egonet is an induced subgraph composed of a central node, its neighbors, and all edges among nodes in the egonet that are also links in the main graph. Therefore, by merging each individual's views of friendship groups, communities can be discovered. The friendship groups represent the small clusters, extracted from egonets, composed of the central node and connected neighbors. More friendship-groups can be combined to create a community. The algorithm consists of two steps; the first one is the detection of friendship groups, while the second step comprises the merging of friendship groups into communities. In step one, the algorithm iterates through every node in the graph, centering on the selected vertex and computing the egonet. After that, friendship groups are extracted from that egonet, and the central node is eliminated, since it is known to exist in multiple friendship groups. Consequently, the graph breaks into multiple connected components. The central vertex is then added back to each connected component obtained to create the friendship groups. The output of the first phase is a set of friendship groups, from an egocentric point of view. The second step consists in merging the groups into communities. This process is done by first merging all exact matches, i.e. groups that are complete or proper subsets of other groups. Finally, groups that are relatively close, i.e.  groups that match all but one element from the smaller group, are merged. The process is repeated until no more merges can be performed.

The same authors, in \cite{Rees2012} presented a \emph{swarm intelligence} approach for overlapping community detection.
A network is considered as a set of agents corresponding to nodes characterized by neighbors.
Each agent interacts with its social groups. The agent knows the set of its friends and even which of their friends are also friends. Friends agree on a common community ID inside the different social groups or friendship groups. The algorithm consists of the following steps. First of all, each agent is assigned an identifier ID. It determines a complete map of its neighborhood and builds an egonet. 
Starting from the egonet, the friendship groups can be extracted by a union-find algorithm. Each friendship group is identified by a unique ID, composed by the base agent ID and a unique incrementing decimal value. Secondly, within each friendship group, the agent will ask to its neighbors their views of the friendship group (which can be various from different perspectives) in order to identify the non-propagating nodes (nodes whose views differ from the agent view and consequently their information is not further propagated). Finally, the assigned friendship groups IDs are propagated. In particular, if the ID value on one of the friendship groups has been modified, the new ID value will be spread to the nodes inside the friendship group. The process is repeated until the convergence in propagation has been reached. At the end of the process, each agent will have a list of assigned communities. Communities can be easily detected because they are groups of agents that share a common ID value.

\subsection{Dynamic Networks} \label{dn} 
The methods described so far do not take into account an important aspect characterizing networks:   i.e. the evolution they go through over time. 
The representation of many complex systems through a static graph, even when the
 temporal dimension describing the varying interconnections among nodes is available,  does not allow to study the network dynamics and the changes it incurs over time.   

\emph{Dynamic networks}, instead, capture the modifications  of interconnections over time, allowing to trace the changes of network structure at different time steps. 
Analyzing networks and their evolution is recently receiving an increasing interest from researchers. However, there have been few proposals for the detection of overlapping communities in dynamic social networks. In the following the more recent methods aiming to seek out dynamic communities are described.  \\

Palla et al. \cite{Palla2007} have been among the first researchers to introduce an approach that allows to analyze the time dependence of overlapping communities on a large scale and as such, to uncover basic relationships characterizing community evolution. Actually they argued that  at each time step communities can be extracted by using the \emph{Clique Percolation Method (CPM)} \cite{Palla05}. The events that characterize the life time of a community are growth or contraction, at each time step a new community can appear, while others can disappear. Furthermore groups can merge or split. In order to identify community evolution along time, the authors proposed to merge networks of two consecutive time steps $t$ and $t+1$, and then apply  the $CPM$ method to extract the new community structure of the joint network. 
Since the joined graph contains the union of the links of the two graphs, any community from time step $t$ to $t+1$ can only grow, merge or remain unchanged, more than one community can be merged into a single community, but no community may loose members. Consequently,  any community in one of the original network can be contained in exactly one community of the joined network. 
Communities in the joint graph provide a way to match communities between time steps $t$  and $t+1$. If a community in the joint graph contains a single community from $t$ and a single community from $t+1$, then they are matched. If the joint group contains more than one community from either time steps, the communities are matched in descending order of their relative node overlap. Overlap is computed for every pair of communities from the two time steps as the fraction of the number of common nodes to the sum of number of nodes of both communities. 
Experiments on two real-life networks showed that large groups remain alive if they undergo dynamic changes. On the contrary, small groups survive if they are stable.\\

Cazabet et al. \cite{Cazabet2010} introduced the concepts of \emph{intrinsic community} and \emph{longitudinal  detection}, and proposed an algorithm named $iLCD$ (\emph{intrinsic Longitudinal Community Detection}) to discover highly overlapping groups of nodes. An intrinsic community is considered one that owns a characteristic deemed meaningful, such as for example being a 4-clique. Longitudinal detection means that, starting from an intrinsic community, new members join gradually like the snowball effect. 
$iLCD$ considers the list of edges ordered with respect to the time they appeared, and, for each edge($u,v$) of the set of edges $E_t$ created at time $t$, it performs three steps.  First, for each community $C$ which $u$ (resp. $v$) belongs to, it tries to add $v$ (resp. $u$), as explained below. Then, if $u$ and  $v$ do not already belong to any community, it tries to create a new one; finally, similar communities are merged.   The first step of updating existing communities by the addition of new nodes is realized by estimating, for each community, the mean number of second neighbors $EMSN$, i.e. nodes that can be reached with a path of length 2 or less, and the mean number of robust second neighbors $EMRSN$, i.e. nodes that can be reached with at least two  paths of length 2 or less. A new node is accepted in the community if the number of its neighbors at rank 2 is greater than $EMSN$, or the number of its robust neighbors at rank 2 is greater than $EMRSN$. The second step of creating a new community checks whether the couple of nodes ($u,v$) constitutes a minimal predefined pattern, like a 4-clique. This intrinsic property must be predetermined. Finally merging is executed by fixing an overlap threshold, and when two communities have an overlap above the threshold, the smaller is deleted and the greater is retained.  This choice, as the authors point out, limits the use of uncertain heuristics. Comparison with other methods shows that this approach outperforms $CPM$ only if the network is highly dense.\\

Another recent proposal to detect overlapping communities in dynamic networks is described in \cite{Nguyen2011} by Nguyen et al. The method, named $AFOCS$ (\emph{Adaptive Finding Overlapping Community Structure}), consists of two phases. In the first phase  local communities are obtained by searching for all the groups of nodes $C$ whose internal density 

\begin{equation}
\Psi(C)=\frac{\mid C^{in} \mid}{\mid C \mid *(\mid C \mid -1)/2}
\end{equation}

\noindent where $C^{in}$ is the number of internal connections of $C$, i.e. the number of links having both endpoints in $C$, is higher than a threshold 
$\tau(C)$ defined as

\begin{equation}
\tau(C)= \frac{\sigma(C)}{\mid C \mid *(\mid C \mid -1)/2}
\end{equation}

\noindent where 

\begin{equation}
\sigma(C) =    (\mid C \mid *(\mid C \mid -1)/2)^{1 - \frac{1}{\mid C \mid *(\mid C \mid -1)/2}}
\end{equation}

\noindent The local communities are then merged provided that their overlapping score is higher than a value given as input parameter. The second phase adaptively updates the communities obtained in the first step, by considering how the network evolves over time. The authors individuate four major changes a network can incur:  a new node and its adjacent edges are either added or removed to/from  the network;  a new edge connecting two existing nodes is added or an existing edge is removed. The algorithm is able to obtain the new community structure by adopting the more apt strategy to determine whether a community will split, or two communities will merge. A comparison with existing approaches showed that $AFOCS$ performances are competitive with other methods, mainly as regards running time.

\begin{table}
\caption{{ A summarization of the reviewed methods. For EAGLE, $h$ is the number of pairs of maximal cliques which are neighbors, and $s$ is the number of maximal cliques; for CGE, $h$ is the number of cliques; for Ahn and AFOCS, $d_{max}$ is the maximum node degree; for GA-NET+ $t$ is the number of generations and $p$ the population size; for SLPA $t$ is the number of iterations performed by the algorithm. }}\label{tab1}
\vspace{1mm}
  \centering
          \footnotesize
\begin{tabular}{p{3cm}p{2cm}p{3cm}p{3cm}}
\hline\noalign{\smallskip}

{\sc Approach}& {\sc Method}& {\sc Reference} & {\sc Complexity}\\
\hline\noalign{\smallskip}

{\sc Node seeds and local expansion}  & {\sc IS, RaRe} &Baumes et al.  &\\ %\cline{2-4}
{\sc } &{\sc $IS^2$}, $CIS$& \cite{Baumes2005b},\cite{Baumes2005a},\cite{Goldberg2010}&$O(mk + n)$\\\cline{2-4}
&$LFM$& Lancichinetti et al. &\\ 
&$OSLOM$&  \cite{Lancichinetti2008a},\cite{Lancichinetti2011}&$O(n^2)$ \\\cline{2-4}
&{\sc DOCS} & Wei et al. \cite{Wei2009}&-\\\cline{2-4}
&{\sc Moses} & McDaid and Hurley \cite{McD2010}&$O(n^2)$\\
\hline\noalign{\smallskip}
{\sc Clique} &{\sc CFinder} &Palla et al. &$O(m^{\frac{ln~m}{10}})$\\
{\sc Expansion} &{\sc CPM} &\cite{Palla06},\cite{Palla05}&\\\cline{2-4}
&{\sc EAGLE}& Shen et al.  \cite{Shen2009}&$O(n^2 +(n+h)s)$\\\cline{2-4}
&{\sc GCE} & Lee et al. \cite{Lee2010}&$O(mh)$\\
\hline\noalign{\smallskip}

{\sc Line Graph}&{\sc Evans} &Evans and Lambiotte \cite{Evans2009},\cite{Evans2010}&$O(2mk~log~n)$\\\cline{2-4}
&{\sc GA-NET+}& Pizzuti \cite{Pizzuti2009} &$O(tp(m+m~log~m)$\\\cline{2-4}
&{\sc Ahn} & Ahn et al.\cite{Ahn2010}&$O(nd^2_{max})$\\
\hline\noalign{\smallskip}

{\sc Label}&{\sc  COPRA} & Gregory \cite{Gregory2010} &$O(vm~log(vm/n))$\\ \cline{2-4}
{\sc Propagation}&{\sc  BMLPA} & Wu et al. \cite{Wu2012} &$O(n~log~n)$\\ \cline{2-4}
& {\sc SLPA} & Xie et al. \cite{Xie2011,Xie2012}&$O(tm)$\\ 
\hline\noalign{\smallskip}

{\sc Dynamic} &{\sc CPM} &Palla et al. \cite{Palla05}&\\\cline{2-4}
{\sc methods} &{\sc iLCD} &Cazabet et al. \cite{Cazabet2010}&$O(nk^2)$\\\cline{2-4}
 &{\sc AFOCS} &Nguyen et al. \cite{Nguyen2011}&$O(d_{max}m)+O(n^2)$\\\cline{2-4}
\hline\noalign{\smallskip}

{\sc Other} &{\sc FCM}&Zhang et al.  \cite{Zhang2007}&$O(nk^2)$\\\cline{2-4}
{\sc methods}&{\sc CONGA} & Gregory \cite{Gregory2007} &$O(m^3)$\\\cline{2-4}
 &{\sc CONGO} & Gregory \cite{Gregory2008} &$O(m log m)$\\\cline{2-4}
&{\sc ONDOCS}&Chen et al. in \cite{Chen2010}&-\\\cline{2-4}
& {\sc Egonet}& Rees and Gallagher \cite{Rees2010} &$O(n(log~n)^2+n^2 ~log~n)$\\\cline{2-4}
& {\sc Swarm Egonet}& Rees and Gallagher \cite{Rees2012} &$O(n~log^2~n)$\\\hline
\hline\noalign{\smallskip}

\end{tabular}
\end{table}

\begin{table}
\caption{{Methods for which it is possible to download the software.}}\label{tab2}
  \centering
          \footnotesize

\begin{tabular}{p{4cm}p{7cm}}
\hline\noalign{\smallskip}

{\sc Reference}  & {\sc Software}\\
\hline\noalign{\smallskip}

 Lancichinetti et al. \cite{Lancichinetti2008a},\cite{Lancichinetti2011}&https://sites.google.com/site/andrealancichinetti/software\\
McDaid and Hurley \cite{McD2010}&https://sites.google.com/site/aaronmcdaid/downloads\\
Palla et al.  \cite{Palla06},\cite{Palla05}&http://www.cfinder.org\\
Lee et al. \cite{Lee2010}&https://sites.google.com/site/greedycliqueexpansion\\
Evans and Lambiotte \cite{Evans2009},\cite{Evans2010}&https://sites.google.com/site/linegraphs/\\
Pizzuti \cite{Pizzuti2009} &http://www.icar.cnr.it/pizzuti/codes.html\\
Ahn et al. \cite{Ahn2010}&https://github.com/bagrow/linkcomm\\
Gregory \cite{Gregory2010} &http://www.cs.bris.ac.uk/~steve/networks/copra/\\ 
Wu et al. \cite{Wu2012}&http://dev.bjtu.edu.cn/bmlpa/\\
Xie et al. \cite{Xie2011,Xie2012}& https://sites.google.com/site/communitydetectionslpa\\ 
Cazabet et al. \cite{Cazabet2010}&http://cazabetremy.fr/Cazabet\_remy/iLCD.html\\
Gregory \cite{Gregory2007} &http://www.cs.bris.ac.uk/~steve/networks/congapaper/\\
Gregory \cite{Gregory2008} &http://www.cs.bris.ac.uk/~steve/networks/congopaper/\\
\hline 
\end{tabular}
\end{table}

\section{Benchmarks for testing algorithms}\label{bench} 
The capability of an algorithm in detecting community structure is usually validated by testing   the method on artificial or real world networks for which the division in communities is known. Since the availability of ground-truth community structure for large real networks is rather difficult, synthetic benchmarks built by specifying parameters to characterize network structure are preferred. 

One of the most known benchmarks for non-overlapping networks has been proposed by Girvan and Newan in \cite{Girvan02}. The network consists of 128 nodes divided into four communities of 32 nodes each. Edges are placed between vertex
pairs at random but such that $z_{in} + z_{out} = 16$, where
$z_{in}$ and $z_{out}$ are the internal and external degree of a
node with respect to its community. If $z_{in} > z_{out}$ the
neighbors of a node inside its group are more than the neighbors
belonging to the other three groups, thus a good algorithm should
discover them.  This benchmark, as observed in \cite{Lancichinetti2009}, however, is rather simple since it is characterized by non-overlapping communities having all the same size and by nodes having the same expected degree. Thus, Lancichinetti et al. \cite{Lancichinetti2008b} proposed a new class of benchmarks that extend the Girvan and Newman's benchmark by introducing power law degree distributions, different community size, and percentage of overlap between communities.  

The benchmark is also characterized by the \emph{mixing parameter} $\mu=\frac{z_{out}}{z_{in} + z_{out}}$ that gives the ratio between the external degree of a node and the total degree of the node. When $\mu < 0.5$ the communities are well defined, thus a good algorithm should discover
them.  The software is public and can be downloaded from http://sites.google.com/site/andrealancichinetti/software.

\section{Discussion and conclusions}\label{concl} 
The paper reviewed state-of-the-art approaches for the detection of overlapped communities. Methods have been classified in different categories. Node seed and local expansion methods together with clique expansion approaches are characterized by the same idea of starting with a node (in the former case), or a group of nodes (in the latter case) and then expanding the current cluster until the adopted quality function increases. Link clustering methods detect overlapping communities by partitioning the set of links rather than the set of nodes by using the \emph{line graph} corresponding to the network under consideration. Since generally the number of edges is much higher than the number of nodes, these methods are computationally expensive. Label propagation approaches are among the most efficient since they start from a node and visit neighboring nodes to propagate class label. Approaches reported in Section \ref{oa}  to find overlapping communities adopt strategies that substantially differ from the others. Thus Zhang et al. \cite{Zhang2007} use modularity, spectral relaxation and fuzzy c-means, Gregory \cite{Gregory2008} relies on the concept of split betweenness to duplicate a node, Chen et al. \cite{Chen2010} propose an interactive approach based on visual data mining, Rees and Gallagher \cite{Rees2010}, \cite{Rees2012} use the concept of egonet and apply swarm intelligence. Dynamic approaches try to deal with the problem of network evolution and constitute a valid help in understanding changes a network might undergo over time. 

A summarization of the  described methods is reported in Table \ref{tab1}. For each method, when known, the computational complexity is reported. Furthermore, in Table \ref{tab2}, a link to the web site from which it is possible to download the software implementing the algorithm is given.

Though the number of approaches present in the literature is notably, the results obtained by each of them are substantially different, thus there is no a universal method that is competitive with respect to all the others for networks having different characteristics such as sparsity, degree distribution, overlap percentage among communities, and so on. As pointed out in \cite{Xie2013},  there are two questions that researchers should focus on: "when to apply overlapping methods and how significant the overlapping is". Investigation on these issues and extensions to weighted networks constitute open problems for future research.

\printglossaries

\end{document}